\documentclass[11pt]{article}
\usepackage{epsfig}
\begin{document}
\title{\bf Distance matrices and isometric embeddings}
\author{E. Bogomolny, O. Bohigas, and \fbox{C. Schmit}\\
Universit\'e Patis-Sud, CNRS, UMR 8626,\\
 Laboratoire de Physique Th\'eorique et Mod\`eles Statistiques,\\
91405 Orsay Cedex, France}

\maketitle

\begin{abstract}
We review the relations between distance matrices and isometric embeddings
and give simple proofs that distance matrices defined on euclidean and 
spherical spaces have all eigenvalues except one non-positive. Several
generalizations are discussed. 
\end{abstract}

\section{Introduction}
Matrices with random (or pseudo-random) elements appear naturally in different
physical problems and their statistical properties  have been thoroughly
investigated (see e.g. \cite{Mehta}).  A special case of random matrices, called distance matrices, has been recently proposed in \cite{Vershik}.  They are  defined for any metric space $X$ with a
probability measure $\mu$ on it as follows. Choose
$N$ points $\vec{x}_j \in X$ randomly distributed according to the measure
$\mu$. The matrix element $M_{ij}$ of the $N\times N$ distance matrix  $M$ 
equals the distance  on $X$ between points
$\vec{x}_i$ and $\vec{x}_j$
\begin{equation}
M_{ij}=\mbox{distance}(\vec{x}_i,\vec{x}_j),\;\; \;\; i,j=1,\ldots,N\ .
\label{mij}
\end{equation}
In \cite{BBC} we have discussed the eigenvalue density for distance matrices defined on certain manifolds.
When first numerical calculations were performed, an intriguing fact was observed, namely, all eigenvalues (except one) of distance matrices on euclidean and spherical manifolds  were non-positive. 
However, this property was not fulfilled e.g. for points on a torus. 
Typically, eigenvalues of generic random matrices occupy the whole available energy space and to impose the condition that all of them but one are non-positive requires the control of the signs of all principal minors (see Section~\ref{schoenberg}) which is usually  difficult to impose. In investigating this fact we have found a direct proof that for distance matrices over manifolds embedded into the euclidean space this property is automatically fulfilled. This relation between very basic geometrical properties of a manifold and spectral properties of its distance matrix was unexpected for us but analysing the literature we found that it has been proved by Schoenberg in the thirties \cite{Schoenberg35, Schoenberg37} and in \cite{BBC} we noted this fact without details. Nevertheless, after many discussions on different occasions it became clear that this type of problems is practically unknown in the physical community and we think that it is of interest to  present simple proofs of the main statements. The material in this note is not new (general  references are \cite{Wells}-\cite{Deza}) but it seems that it has not been discussed in the random matrix community.

By definition of distance  the matrix elements of a distance matrix have the following
properties\\
a) positivity
  \begin{equation} 
    M_{ij}\geq 0\;\;\mbox{and $M_{ij}=0$ only when $i=j$}\ ,
  \label{positivity}
  \end{equation}
b) symmetry
  \begin{equation}
    M_{ij}=M_{ji}\ ,
  \label{symmetry}
  \end{equation}
c) triangular inequalities
  \begin{equation}
M_{ij}\le M_{ik}+M_{kj}\;\;\;\;\mbox{for all $i$, $j$, $k$}\ .
  \label{triangular}
  \end{equation}
Eigenvalues, $\Lambda_p$, and eigenvectors, $u^{(p)}_j$, of distance
matrices are defined in the usual way
\begin{equation}
\sum_{j=1}^{N}M_{ij}u_j^{(p)}=\Lambda_p u_{i}^{(p)}\ .
\label{eigenvalue}
\end{equation}
Distance matrices (\ref{mij}) are real symmetric matrices and 
their eigenvalues are real. As all matrix elements of distance matrices are
non-negative, the application of the Perron-Frobenius theorem
(\cite{Gantmacher} V. 2 p. 49) states that these matrices have one special 
positive eigenvalue $\Lambda_0>0$ with the largest modulus. All other eigenvalues
obey the inequality
\begin{equation}
|\Lambda_j|\leq \Lambda_0\ .
\end{equation}
As distance matrices have only real eigenvalues the equality is possible only if there is a 
negative eigenvalue $\Lambda'=-\Lambda_0$. 

The subject  of this note is to demonstrate  that eigenvalues of
distance matrices defined on the euclidean  or a spherical
space\footnote{It means that the points $\vec{x}_j$ lie in the $d$-dimensional euclidean space or on a  
sphere.}  are all non-positive 
\begin{equation}
\Lambda_i\leq 0,\;\;\;\;i=1,\ldots,N-1\ ,
\label{main}
\end{equation}
except the above-mentioned Perron-Frobenius eigenvalue $\Lambda_0$ and that this
remarkable property is mainly a consequence  of the possibility of isometric
embedding of a finite metric space with a given distance matrix into the
euclidean space. 

We also remark that if, instead of the distance matrix (\ref{mij}), one
considers new matrices 
\begin{equation}
M_{ij}^{(\gamma)}=[\mbox{distance}(\vec{x}_i,\vec{x}_j)]^{\gamma}\ ,
\end{equation}
their eigenvalues also obey inequality (\ref{main}) provided the exponent in the range 
$0<\gamma \leq 2$ for the euclidean space and $0<\gamma \leq 1$ for the
spherical one.

The plan of this paper is the following. In Section~\ref{iso} it is demonstrated that  
if a finite metric space can be
isometrically embedded into the euclidean space, then the matrix whose
elements are the squares of  distances between initial points is of
 negative type (cf. (\ref{ineq})-(\ref{rest})). The inverse theorem is also true, namely, if a matrix $N$ is 
of  negative type then the matrix with elements $\sqrt{N_{ij}}$ can be embedded into the euclidean space. 
A direct proof of the main
theorem that matrices of negative type have all eigenvalues, except one,
non-positive is presented in Section \ref{schoenberg}. 
In Section \ref{transf}
it is demonstrated that if a matrix $N_{ij}$ is of negative type, a new
matrix $N_{ij}^{\gamma}$ with $0<\gamma\leq 1$ will also be of negative type. 
The general form of such metric transforms is also shortly discussed in this Section.
In Section~\ref{spherical}  spherical spaces are discussed and in 
Section~\ref{embedding} a simple proof that geodesic distance matrices for the
spherical spaces are of negative type is presented. A resum\'e of the results is
given in Section \ref{conclusion}. The derivation
of the Cayley-Menger formula for the volume of a multi-dimensional simplex is
reproduced for completeness in the Appendix.  

\section{Isometric embedding}\label{iso}

Assume that we know a finite matrix $M$ whose 
matrix elements $M_{ij}$ ($i,j=1,\ldots ,N$) obey all properties of a
distance (\ref{positivity})-(\ref{triangular}). The isometric
embedding into the euclidean space consists in finding points $\vec{x}_i$, if any,  belonging to an
euclidean space $R^n$ such that the euclidean distance between each pair of
points $i, j$ coincides with $M_{ij}$
\begin{equation}
||\vec{x}_i-\vec{x}_j||=M_{ij}\ ,
\label{isometric}
\end{equation}
for all $i,j=1,\ldots ,N$. Here $||\ldots ||$ is the euclidean distance
\begin{equation}
D_{ij}\equiv ||\vec{x}_i-\vec{x}_j||=\sqrt{\sum_{k=1}^n\left (x_{i}^{(k)}-x_{j}^{(k)}\right )^2}
\label{euclidean}
\end{equation}
and $x_{i}^{(k)}$ with $k=1,\ldots ,n$ are the
euclidean coordinates of the $n$-dimensional point $\vec{x}_i$. 

The necessary and sufficient conditions of the existence of solutions of
Eq.~(\ref{isometric}) can be obtained from the following considerations
\cite{Schoenberg35} and \cite{Wells, Blumenthal}. 
Choose a point, say $\vec{x}_N$ and consider the vectors
$\vec{y}_i=\vec{x}_i-\vec{x}_N$ with $i=1,\ldots ,N-1$. They form a simplex
in the $n$-dimensional space $R^{n}$. Construct the $(N-1)\times n$  
matrix of coordinates of these vectors
\begin{equation}
V_{ik}=y_i^{(k)}\;\;\;i=1,\ldots ,N-1,\;\;\;k=1,\ldots ,n
\end{equation}
and multiply it by its  transpose. The result is a $(N-1)\times (N-1)$
real symmetric matrix $C=V^T\cdot V$ of scalar products
\begin{equation}
 C_{ij}=\vec{y}_i\cdot \vec{y}_j,\;\;\;i,j=1,\ldots ,N-1\ .
\end{equation}
Because vectors $\vec{y}_j$ belong to the euclidean space their scalar
products can be expressed through the distances between points
\begin{equation}
\vec{y}_i\cdot\vec{y}_j=
\frac{1}{2}(||\vec{y}_i||^2+||\vec{y}_j||^2-||\vec{y}_i-\vec{y}_j||^2)\ .
\label{scalarproduct}
\end{equation}
Therefore the matrix $C_{ij}$ can be calculated from the squares of matrix 
elements of the distance matrix
\begin{equation}
C_{ij}=\frac{1}{2}(M_{iN}^2+M_{jN}^2-M_{ij}^2)\ .
\label{embed}
\end{equation}
If points $\vec{x}_j$ obeying Eqs.~(\ref{isometric}) do exist then by
construction the matrix $C_{ij}$ is such that the quadratic form 
\begin{equation}
(\xi C \xi)=\sum_{i,j=1}^{N-1}C_{ij}\xi_i\xi_j\equiv \left (\sum_{i=1}^N\xi_i\vec{y}_i\right )^2\  
\label{form}
\end{equation}
is non-negative $(\geq 0)$ for any choice of real numbers
$\xi_1,\xi_2,\ldots ,\xi_{N-1}$. Inversely, if one has a symmetric
positive matrix $C$, it can be written in the form
\begin{equation}
C=U^TU,
\end{equation}
where the matrix $U$ can be chosen, e.g., in the lower  triangular
form (the Cholesky decomposition). Then the elements of $U=y_i^{(k)}$ give
directly the coordinates $y_i^{(k)}$ of points obeying (\ref{embed}) which
solve the problem of  the isometric embedding.

The quadratic form (\ref{form}) can be rewritten in a simpler form by
introducing  a new variable $\xi_N=-\sum_{j=1}^{N-1}\xi_j$. Then
\begin{equation}
(\xi C \xi)=-\frac{1}{2}\sum_{i,j=1}^N M_{ij}^2\xi_i\xi_j\ .
\end{equation}
Therefore the necessary and sufficient condition of the existence of isometric embedding of
a finite metric space with the distance matrix $M$ into the  euclidean space  is that a new matrix $N$ whose matrix elements equal the square of matrix elements of the matrix $M$
\begin{equation}
 N_{ij}=M_{ij}^2
\label{existence}
\end{equation}
is such that the quadratic form associated with it
\begin{equation}
(\xi N \xi)=\sum_{i,j=1}^N N_{ij} \xi_i\xi_j
\label{condition}
\end{equation}
is non-positive 
\begin{equation}
\sum_{i,j=1}^N \xi_i N_{ij}\xi_j \leq 0
\label{ineq}
\end{equation}
for all choices of real numbers $\xi_j$, $j=1,\ldots ,N$ with zero sum
\begin{equation}
\sum_{j=1}^N\xi_j=0\ .
\label{rest}
\end{equation}
In general, a real symmetric  matrix obeying these conditions is called a matrix of negative
type.

The Schoenberg theorem states that if a metric space with a
distance matrix $M_{ij}$ can be isometrically embedded into the euclidean
space, the matrix $M_{ij}^2$ is of negative type and if a matrix $N_{ij}$ is of negative type, the metric space with
the distance matrix $\sqrt{N_{ij}}$ can be isometrically embedded into the
euclidean space. The minimal dimension of the embedded euclidean space is the 
rank of the matrix  $C_{ij}$ in Eq.~(\ref{embed}). 

\section{Eigenvalues of negative type matrices}\label{schoenberg}

In this Section we present, following \cite{Schoenberg37}, the direct proof
that any  matrix $N_{ij}$ of the negative type has all eigenvalues except one non-positive ($\leq 0$). An indirect proof of this statement can be found in \cite{Deza}.

The law of inertia (see e.g. \cite{Gantmacher} V. 1 p.298) states that if a
real quadratic form
\begin{equation}
(x N x)=\sum_{i,j=1}^N N_{ij}x_ix_j
\end{equation}
is transformed into a sum of squares of independent linear forms
$X_i=\sum_{j=1}^Nc_{ij}x_j$
\begin{equation}
(x N x)=\sum_{i=1}^{r}b_iX_i^2
\label{inertia}
\end{equation}
then the total number of positive and negative coefficients $b_i$ is
independent of the representation.
In particular, in the eigenbasis of the real symmetric matrix $N_{ij}$ 
\begin{equation}
(x N x)=\sum_{i=1}^N \Lambda_i u_i^2
\end{equation}
and the law of inertia permits to determine the number of positive and
negative eigenvalues $\Lambda_i$.

According to the Jacobi theorem (see e.g. \cite{Gantmacher} V. 1 p. 305) if
the principal minors $\Delta_j$ of a matrix  are non-zero then the
number of positive (resp. negative) terms in (\ref{inertia}) coincides
with the number of conservation (resp. alteration) of signs in the sequence
\begin{equation}
1,\Delta_1,\Delta_2,\ldots,\Delta_N 
\label{jacobi}
\end{equation}
(we assume that the matrix $a_{ij}$ is of full rank).
Recall that 
the principal minor $\Delta_n$ of a matrix $N$ is the determinant of the
left-upper $n\times n$ sub-matrix
\begin{equation}
\Delta_n=\left | \begin{array}{ccccc}
N_{11}&N_{12}&N_{13}&\ldots&N_{1n}\\  
N_{12}&N_{22}&N_{23}&\ldots&N_{2n}\\
\vdots&\vdots&\vdots&\vdots&\vdots\\
N_{1n}&N_{2n}&\ldots&N_{n-1\ n}&N_{nn}
\end{array} \right |\ .
\label{principalminor}
\end{equation}
For distance matrices $N_{ii}=0$ and $\Delta_1\equiv 0$ which prevents the
direct application of the Jacobi theorem. This formal difficulty
can, for example, be overcome as follows. It is clear that eigenvalues and
other principal minors of generic distance matrices are non-zero. Therefore
if  one adds to $N$ a diagonal
matrix $\epsilon \delta_{ij}$ with small $\epsilon$ the signs of eigenvalues
will not change. But in such a case $\Delta_1=\epsilon$ and the sequence
(\ref{jacobi}) takes the form
\begin{equation}
1,\epsilon,\Delta_2,\Delta_3,\ldots, \Delta_N\ .
\label{seq2}
\end{equation}
We shall prove below that principal minors $\Delta_n$ of distance matrices 
of the negative type  have alternating sign
\begin{equation}
\Delta_n=(-1)^{n-1} v_n^2,\;\;n=2,3,\ldots ,N\ .
\label{sign}
\end{equation}
Irrespective of the sign of $\epsilon$  there is one conservation of sign and
$N-1$ alterations of  signs in the sequence (\ref{seq2}). Therefore, 
according to the Jacobi theorem, distance  matrices of the negative type 
have one positive (the Perron-Frobenius) eigenvalue and all other eigenvalues 
are non-positive. 

Because the matrix $N_{ij}$ is of the negative type, the metric space 
with the distance $\sqrt{ N_{ij}}$ can be embedded into the euclidean space. It
means that there exist points $\vec{x}_j$ in the euclidean space such that
the euclidean distances between any pairs of points equal 
\begin{equation}
\tilde{D}_{ij}=\sqrt{N_{ij}}\ .
\end{equation}
Let us consider one of these points as the origin (say $\vec{x}_1$).
Points $\vec{x}_2,\ldots ,\vec{x}_n$ can be viewed as vertices of a 
$(n-1)$-dimensional simplex. Denote $\tilde{D}_{1i}=r_i$. Then the distance between
any pair of points can be expressed as follows
\begin{equation}
\tilde{D}_{ij}=\sqrt{r_i^2+r_j^2-2r_ir_j\cos \varphi_{ij}}\ ,
\end{equation}
where $\varphi_{ij}$ is the euclidean angle between vectors $\vec{x}_i-\vec{x}_1$ and
$\vec{x}_j-\vec{x}_1$. 

Let us perform an inversion $r_i\to  1/r_i$ for all $i=2,\ldots ,n$.
Then instead of $n-1$ points $\vec{x}_2,\ldots ,\vec{x}_n$  we get a new set of
$n-1$ euclidean points $\tilde{\vec{x}}_2,\ldots ,\tilde{\vec{x}}_n$ 
whose mutual distances $D_{ij}$ can be expressed through the old 
distances as 
\begin{equation}
D_{ij}=
\sqrt{\frac{1}{r_i^2}+\frac{1}{r_j^2}-2\frac{1}{r_ir_j}\cos \varphi_{ij}}=
\frac{\tilde{D}_{ij}}{\tilde{D}_{1i}\tilde{D}_{1j}}\ .
\label{transform}
\end{equation}
Because the points $\vec{x}_j$ belong to the euclidean space the new points
$\tilde{\vec{x}}_j$ with $j=2,\ldots , N$ plus the point $\vec{x}_1$ form
a $n-1$-dimensional euclidean simplex. The volume
of this simplex can be computed by  the Cayley-Menger
determinantal formula (see e.g. \cite{Sommerville} p.124 and also \cite{Berlin} for an early reference) which expresses
the volume  $V(P_1,\ldots ,P_n)$ of a $n$-dimensional euclidean simplex 
through the lengths of its sides
\begin{equation}
V^2(P_1,\ldots ,P_n)=\frac{(-1)^n}{2^{n-1}[(n-1)!]^2}D(P_1,\ldots ,P_n)\ ,
\label{volume}
\end{equation}
where the Cayley-Menger determinant is
\begin{equation}
D(P_1,\ldots ,P_n)=\left |\begin{array}{ccccc}
0&1&1&\ldots &1\\
1&0&D_{12}^2&\ldots &D_{1n}^2\\
\vdots&\vdots&\vdots&\vdots&\vdots\\
1&D_{1n}^2&D_{2n}^2&\ldots &0
\end{array}\right |\ ,
\label{CM}
\end{equation}
and $D_{ij}$ are the distances between points $i$ and $j$ for $i,j=1,\ldots ,n$.  
For completeness we give in Appendix a derivation of this formula.

In our case the lengths of the transformed simplex are given by
Eq.~(\ref{transform}). As each $\tilde{D}_{ij}=\sqrt{N_{ij}}$, the squares of the
lengths which enter the Cayley-Menger formula (\ref{CM}) are
\begin{equation}
D_{ij}^2=\frac{N_{i,j}}{N_{1i}N_{1j}}\ .
\end{equation}
Therefore for each $n=2,\ldots ,N$
\begin{equation}
D(\vec{y}_2,\ldots \vec{y}_n)\equiv 
\left |\begin{array}{ccccc}
0&1&1&\ldots &1\\
1&0&\frac{N_{23}}{N_{12}N_{13}}&\ldots &\frac{N_{2n}}{N_{12}N_{1n}}\\
1&\frac{N_{32}}{N_{13}N_{12}}&0&\ldots &\frac{N_{3n}}{N_{13}N_{1n}}\\
\vdots&\vdots&\vdots&\vdots&\vdots\\
1&\frac{N_{n2}}{N_{1n}N_{12}}&\frac{N_{n3}}{N_{1n}N_{1n}}&\ldots &0
\end{array}\right |\ .
\end{equation}
As the determinant is a multi-linear form of row and columns by multiplication of each row $(ij)$ and 
each column $(ji)$ by $N_{ij}$ one gets
\begin{equation}
D(\vec{y}_2,\ldots \vec{y}_n)= [N_{12}N_{13}\ldots N_{1N}]^{-2}
\left |\begin{array}{ccccc}
0&N_{12}&N_{13}&\ldots &N_{1n}\\
N_{21}&0&N_{23}&\ldots &N_{2n}\\
N_{31}&N_{32}&0&\ldots &N_{3n}\\
\vdots&\vdots&\vdots&\vdots&\vdots\\
N_{n1}&N_{n2}&N_{n3}&\ldots &0
\end{array}\right |\ .
\end{equation}
But the determinant in this expression coincides with the principal minor of
the initial distance matrix. Therefore
\begin{equation}
\Delta_n=(-1)^{n-1} [N_{12}N_{13}\ldots N_{1N}]^{2}2^{n-1}[(n-1)!]^2
V^2(\vec{y}_1,\ldots ,\vec{y}_n)\ ,
\end{equation}
which  proves that the
principal minors of matrices of the negative type are of alternate
signs.   This relation, as  explained above, implies that
all eigenvalues of such matrices (except possibly one) are non-positive. 

\section{Metric transform}\label{transf}

The problem of isometric embedding gives rise to different
generalizations. One type of question is the following. Let the points $\vec{x}_j$ 
with $j=1,\ldots ,N$ be  points of the euclidean space $R^n$. Find all functions
$F(r)$ (called  metric transforms) such that the finite metric space with the
distance matrix 
\begin{equation}
  M_{ij}=F(||\vec{x}_i-\vec{x}_j||)
\end{equation}
can be embedded into an euclidean space $R^k$ with certain $k$. Here
$||\vec{x}_i-\vec{x}_j||$ is the euclidean distance (\ref{euclidean}) 
between point $\vec{x}_i$ and $\vec{x}_j$. 

In \cite{Schoenberg37} it was proved that general metric transforms can be
expressed through radial positive definite functions. A real function
$f(r )$ is called radial positive definite  provided 
\begin{equation}
\sum_{i,j=1}^Nf(||\vec{x}_i-\vec{x}_j||)\xi_i \xi_j\geq 0
\end{equation}
for all choices of points $\vec{x}_i\in R^n$ and of  real numbers $\xi$. 

An important example of such a function is
\begin{equation}
f(r)=\exp (-\lambda^2 r^2)\ .
\end{equation}
The positive definite property of this function is the direct consequence of
the well known formula
\begin{equation}
\exp (-\lambda^2||\vec{x}||^2)=\frac{1}{(4\pi)^{n/2}}\int_{R^n}
  e^{i\lambda \vec{x}\cdot \vec{k}}\exp (-||\vec{k}||^2)d^n k\ ,
\end{equation}
from which it follows that 
\begin{equation}
\sum_{i,j=1}^N\xi_i\xi_je^{-\lambda^2 ||\vec{x}_i-\vec{x}_j||^2}=
\frac{1}{(4\pi)^{n/2}}\int_{R^n}\left |\sum_{i=1}^N \xi_i 
e^{i\lambda \vec{x}_i\cdot \vec{k}}\right |^2 e^{-||\vec{k}||^2}d^n k \geq 0\ .
\end{equation}
The following theorem is easily proved \cite{Schoenberg37}. The finite metric
space with a distance matrix $M_{ij}$ can be isometrically
embedded into the euclidean space if and only if  the quadratic form 
\begin{equation}
\sum_{i,j=1}^N\exp (-\lambda^2 M_{ij}^2)\xi_i\xi_j
\label{expM2}
\end{equation}
is non-negative  ($\geq 0$)
for all choices of real numbers $\xi_j$ and  all $\lambda \to 0$. 

The proof
is as follows. If the space can be isometrically embedded into the euclidean space then there exist
points
$\vec{x}_j\in R^n$ such that $M_{ij}=||\vec{x}_i-\vec{x}_j||$. Because 
$\exp(-\lambda^2 r^2)$ is a radial positive definite function  the
quadratic form (\ref{expM2}) is non-negative. Conversely, if the quadratic form is
non-negative for $\lambda\to 0$ then 
\begin{equation}
 \sum_{i.j=1}^N (1-\lambda M_{ij}^2)\xi_i\xi_j\geq 0
\end{equation}
for all $\xi_j$. Choosing $\sum_{j=1}^N\xi_j=0$ cancels the first term and
reduces the above inequality to (\ref{condition}), thus proving the
existence of the embedding.

The fact that $\exp (-\lambda^2 r^2)$ is a radially positive definite function
permits also  to prove \cite{Schoenberg37} that the metric space with the 
distance equal a power of the euclidean distance
\begin{equation}
M_{ij}'=||\vec{x}_i-\vec{x}_j||^{\gamma}\;\;\;i,j=1,\ldots ,N\ ,
\label{gamma}
\end{equation}
where $\vec{x}_i\in R^n$ and $0<\gamma\leq 1$ can be embedded into the euclidean
space. The proof follows from the identity valid for 
$0<\gamma \leq 1$ 
\begin{equation}
|t|^{2\gamma}=c_{\gamma}\int_0^{\infty}(1-\exp (-\lambda^2 t^2))
\frac{d\lambda}{\lambda^{1+2\gamma}}\ ,
\end{equation}
with
\begin{equation}
c_{\gamma}^{-1}=\int_0^{\infty} (1-\exp( \lambda^2))
\frac{d\lambda}{\lambda^{1+2\gamma}}>0\ .
\end{equation}
One has
\begin{eqnarray}
 \sum_{ij}||\vec{x}_i-\vec{x}_j||^{2\gamma} \xi_i\xi_j =c_{\gamma}\sum_{ij}
 \xi_i\xi_j 
 \int_0^{\infty}(1-e^{-\lambda^2||\vec{x}_i-\vec{x}_j||^2})
 \frac{d\lambda}{\lambda^{1+2\gamma}}\nonumber\\
=c_{\gamma}
\int_0^{\infty}[(\sum_{i}\xi_i)^2-
(\sum_{ij}e^{-\lambda^2||\vec{x}_i-\vec{x}_j||^2}\xi_i\xi_j)]
\frac{d\lambda}{\lambda^{1+2\gamma}}\ .
\end{eqnarray}
If $\sum_{i}\xi_i=0$ the first term is zero and as $e^{ -\lambda^2 r^2}$ is 
radial positive definite, the right-hand side is negative which proves that the
matrix 
\begin{equation}
||\vec{x}_i-\vec{x}_j||^{2\gamma}
\end{equation}
with $0<\gamma \leq 1$ is of negative type and the metric space with
the distance (\ref{gamma}) can be embedded into the euclidean space. 

Combining together the above theorems, one concludes that if a matrix
$N_{ij}$ is of negative type, then the matrix $N_{ij}^{\gamma}$ with
$0<\gamma \leq 1$ is also of negative type and all its eigenvalues, except
at most one, are non-positive. 

General radial positive definite functions $f(r)$ have the form  \cite{Schoenberg37}
\begin{equation}
f(r)=\int_0^{\infty} \Omega_N(ru)d\mu (u)\ ,
\end{equation}
where the measure $\mu$ is non-negative, $\mu(u)\geq 0$, and the function $\Omega_N(r)$
is the integral of $e^{i\vec{k}\cdot \vec{x}}$ with $||\vec{x}||=r$ 
over the $(N-1)$-dimensional sphere 
\begin{equation}
\Omega_N(r)=\frac{1}{\omega_{N-1}}
\int_{S_{N-1}}e^{i\vec{x}\cdot \vec{k}}d\sigma_{N-1}=
\Gamma \left (\frac{N}{2}\right )\left (\frac{2}{r}\right )^{(N-2)/2}J_{(N-2)/2}(r)\ .
\end{equation}
Here $\omega_{N-1}=2\pi^{N/2}/\Gamma (N/2)$ is the volume of
the $(N-1)$-dimensional sphere, $\Gamma (x)$ is the Gamma function, and
$J_n(x)$ is the Bessel function.

From the theorem (\ref{expM2}) it follows \cite{Schoenberg37} that the 
general form of a metric transform is
\begin{equation}
F(r)=\left \{\int_0^{\infty}
\frac{1-\Omega_N(ru)}{u^2}d\nu(u)\right \}^{1/2}
\label{formula}
\end{equation}
with a positive measure $\nu(u)$ such that $\int_0^{\infty}d \nu(u)/u^2$
exists. 

\section{Spherical spaces}\label{spherical}

Eq.~(\ref{formula}) gives the general form of the metric transforms which
transform an euclidean space into another euclidean space. Similar questions
can be asked
about the unit radius  spherical spaces\footnote{Modifications for spherical spaces of radius $R$
  are evident} $S_{d-1}$  which consist of  points
$\vec{x}_j\in R^d$ obeying
\begin{equation}
  \vec{x}_1^2+\vec{x}_2^2+\ldots +\vec{x}_d^2=1\ .
\end{equation}
The geodesic distance on the sphere is 
\begin{equation}
d(\vec{x},\vec{y}\ )=\arccos (\vec{x}\cdot \vec{y}\ )\ .
\label{sphericaldistance}
\end{equation}
The necessary and sufficient conditions that a metric space with the distance
matrix $M_{ij}$ can be embedded isometrically into the spherical space with the distance 
(\ref{sphericaldistance}) coincide with the condition that $N$ initial points plus one point at the origin 
can be  embedded into the euclidean space. From (\ref{scalarproduct}) it follows that
the later can be expressed as the non-negativity  condition of the quadratic form
\begin{equation}
\sum_{i,j=1}^N\cos (M_{ij}) \xi_i\xi_j\geq 0
\end{equation}
for all choices of real numbers $\xi_j$.

Similarly as for the euclidean spaces one can find all positive definite functions
on the spherical spaces. In \cite{Schoenberg42} it was proved that these
functions have the form
\begin{equation}
g(t)=\sum_{l=0}^{\infty} a_l C_l^{p/2}(\cos t)\ ,
\label{positive}
\end{equation}
where all coefficients $a_l$ are non-negative $a_l\geq 0$.
Here $p=d-2$ and $C_l^k(\cos t)$ are the Gegenbauer polynomials. 

This condition can easily be understood from the expression of the 
Gegenbauer polynomial through the orthogonal set 
of the hyper-spherical harmonics $Y_l^{(k)}(\vec{x}\ )$ (see e.g. \cite{Bateman}, 11.4.2) 
\begin{equation}
 \frac{C_l^{p/2}(\vec{x}\cdot \vec{y}\ )}{C_l^{p/2}(1)}=\frac{\omega_{d-1}}{h(p,l)}
\sum_{k=1}^{h(p,l)}Y_l^{(k)}(\vec{x}\ )Y_l^{(k)}(\vec{y}\ )\ ,
\label{Gegenbauer}
\end{equation}
where $h(p,l)$ is the dimension of the irreducible representations of the
$d-1$ dimensional rotation group
\begin{equation}
h(p,l)=(2l+p)\frac{(l+p-1)!}{p!\ l!}\ .
\end{equation}
If Eq.~(\ref{positive}) is fulfilled, one has
\begin{equation}
\sum_{i,j=1}^Ng(d(\vec{x}_i,\vec{x}_j))\xi_i\xi_j=
\omega_{d-1}\sum_{l=0}^{\infty} \frac{a_l}{h(p,l)} 
\sum_{k=1}^{h(p,l)}\left |\sum_{j=1}^N Y_l^{(k)}(\vec{x}_j)\xi_j \right |^2\ ,
\end{equation}
which is evidently non-negative ($\geq 0$).

\section{Embedding of the spherical space into the euclidean
  space}\label{embedding}

In this Section we show  that distance matrices resulting from  spherical geodesic distances are of negative type and,
consequently, the metric space with the distance equal the square root of 
spherical distances can be embedded into the euclidean space.

The proof is based on a following lemma:  the spherical geodesic distance
(\ref{sphericaldistance}) has the expansion
\begin{equation}
d(\vec{x},\vec{y})\equiv \arccos (\vec{x}\cdot \vec{y}\, )=
\lambda_0+
\sum_{l={\mbox{{\scriptsize odd}}}}\lambda_l C_l^{p/2}(\vec{x}\cdot \vec{y})\ ,
\label{expan}
\end{equation}
where all $\lambda_l$ with odd $l$  are negative but  $\lambda_0$ is positive.

Eq.~(\ref{expan}) is the expansion of 
$\arccos (t)$ over   $d$-dimensional spherical harmonics. The coefficients
$\lambda_l$ of this series are 
\begin{equation}
\lambda_l=\frac{1}{h_l(p)}  
\int_0^{\pi}\theta C_l^{p/2}(\cos \theta)\sin^p\theta d\theta\ ,
\end{equation}
where $h_l(p)$ is the normalization integral of the Gegenbauer polynomials
\begin{equation}  
h_l(p)=\int_0^\pi [C_l^{p/2}(\cos \theta)]^2 \sin^{p}\theta d\theta 
\end{equation}
whose explicit expression is (see e.g. \cite{Bateman}, 10.9.7)
\begin{equation}
h_l(p)=\frac{\sqrt{\pi}(l+p-1)!\Gamma ((p+1)/2)}{(l+p/2)l!(p-1)!\Gamma (p/2)}\ .
\end{equation}
As $C_0^{\lambda}=1$, one gets 
\begin{equation}
\lambda_0=\frac{\pi}{2}\ .
\end{equation}
To compute $\lambda_l$ with $l\neq 0$ it is convenient to use 
the Gegenbauer integral (see e.g. \cite{Bateman},  10.9.38)
\begin{equation}
n!\int_0^{\pi} e^{iz\cos \theta }C_n^{\lambda}(\cos
\theta)\sin^{2\lambda}\theta d\theta =2^\lambda \sqrt{\pi} 
\Gamma (\lambda+1/2) \frac{\Gamma(n+2\lambda)}{\Gamma (2\lambda)}i^n
z^{-\lambda} J_{n+\lambda}(z)\ ,
\end{equation}
from which one obtains (cf. \cite{Bateman}, 11.4)
\begin{equation}
  \lambda_l=i^l 2^{p/2}(l+p/2)\Gamma (p/2)\int_{-\infty}^{\infty}
  t^{-p/2}J_{l+p/2}(t)\hat{f}(t)dt\ ,
\end{equation}
where $\hat{f}(t)$ is the Fourier transform of the initial function 
\begin{equation}
\hat{f}(t)=\frac{1}{2\pi} \int_0^{\pi} \theta \sin \theta e^{-i t \cos \theta}
d\theta=\frac{1}{2it} (e^{it}-J_0(t))\ .
\end{equation}
Corresponding to the two terms in $\hat{f}(t)$ there are two terms in $\lambda_l$.
The integral including $e^{it}$ is zero for all $l\neq 0$ 
and the integral with $J_0(t)$ is zero for even $l$.
For  odd $l$
\begin{equation}
  \lambda_l=-i^{l-1}2^{p/2}(l+p/2)\Gamma (p/2)\int_0^{\infty}
  t^{-1-p/2}J_{l+p/2}(t)J_0(t)dt\ .
\end{equation}
The last integral can be computed using the integral 
(\cite{Bateman}, 7.7.4.30)
\begin{eqnarray}
&&\int_0^{\infty} t^{-\rho}J_{\mu}(t)J_{\nu}(t)dt=\\
&&\frac{\Gamma (\rho)\Gamma ((1+\nu+\mu-\rho)/2)}{2^{\rho}
  \Gamma ((1+\nu-\mu+\rho)/2) \Gamma ((1+\nu+\mu+\rho)/2)
  \Gamma ((1+\mu-\nu+\rho)/2)}\ .
\nonumber
\end{eqnarray}
The final result is
\begin{equation}
\lambda_l=-\frac{p(p+2l)}{8\pi}\left [\frac{\Gamma (p/2)\Gamma (l/2)}{\Gamma
  (1+(l+p)/2)}\right ]^2\ .
\end{equation}
This expression is negative which proves the lemma.

Using this lemma and Eq.~(\ref{Gegenbauer}), one concludes that
\begin{equation}
\sum_{i,j=1}^N d(\vec{x}_i,\vec{x}_j)\xi_i\xi_j =\lambda_0(\sum_{i=1}^N \xi_i)^2+
\sum_{l=\mbox{{\scriptsize odd}}} \frac{\lambda_l}{h(p,l)}\sum_{k=1}^{h(p,l)}
\left |\sum_{j=1}^N \xi_j Y_l^{(k)}(\vec{x}_j))\right |^2\ .
\end{equation}
As all $\lambda_l$ with $l\geq 1$ are negative, this expression is negative
for all choices of $\xi_j$ such that $\sum_{j=1}^N\xi_j=0$, i.e. the
spherical geodesic distance matrices are of negative type.

From the theorem of the preceding Sections it follows that 
a new metric space with the distance
\begin{equation}
d^{(\gamma)}(\vec{x},\vec{y})=[\arccos (\vec{x}\cdot\vec{y}\ )]^{\gamma}
\end{equation}
is also of negative type when $0<\gamma \leq 1$ and the space with the
distance 
\begin{equation}
[\arccos (\vec{x}\cdot \vec{y}\ )]^{\gamma/2}
\end{equation}
can be isometrically embedded into the euclidean 
space.

\section{Conclusion}\label{conclusion}

The distance matrices for points in the euclidean and spherical spaces are
of negative type and, consequently, they have all eigenvalues, except one,
non-positive. 

More generally, if points $\vec{x}_j$ belong to the euclidean space, the
above statement is true for the matrices
\begin{equation}
||\vec{x}_i-\vec{x}_j||^{2\gamma}
\end{equation}
with $0<\gamma\leq 1$.

If points $\vec{x}_j$ belong to the spherical space with the distance
$d(\vec{x}_i,\vec{x}_j)$ given by 
Eq.~(\ref{sphericaldistance}) then the matrix 
\begin{equation}
d^{\gamma}(\vec{x}_i,\vec{x}_j)
\end{equation}
with $0<\gamma\leq 1$ is of negative type and has all eigenvalues, except
one, non-negative.

The following theorems are also of interest.

The matrices with elements
\begin{equation}
\exp (-\lambda^2 ||\vec{x}_i-\vec{x}_j||^{2\gamma})
\end{equation}
with $0<\gamma \leq 1$ are positive definite and have all eigenvalues
positive for all $\lambda\to 0$. For $\gamma=1$ this
fact has been mentioned in \cite{Mezard}.

The similar theorem for the spherical space states that matrices 
\begin{equation}
\exp (-\lambda^2 d^{\gamma}(\vec{x}_i,\vec{x}_j))
\end{equation}
with $0<\gamma \leq 1$ are positive definite for all $\lambda$.

\section*{Acknowledgments}

The authors are indebted to  A.M. Vershik for discussion on his work
\cite{Vershik} prior to publication and to L. Pastur for stimulating remarks.

\section*{Appendix}\label{cayley}

The purpose of this Appendix is to give, following \cite{Sommerville}, a
proof of the Cayley-Menger determinantal formula (\ref{volume}).

The volume $V_n$ of the $n$-dimensional Euclidean simplex with one vertex on a point
$\vec{x}_{n+1}$ and $n$ vertices on points $\vec{x}_j$ with $j=1,\ldots ,n$
is proportional to the determinant of components of the $n$ vectors
$\vec{x}_j-\vec{x}_{n+1}$ 
\begin{equation}
V_n=\frac{1}{n!}
\left |\begin{array}{cccc}
x_1^{(1)}-x_{n+1}^{(1)}&x_1^{(2)}-x_{n+1}^{(2)}&\ldots&x_1^{(n)}-x_{n+1}^{(n)}\\
x_2^{(1)}-x_{n+1}^{(1)}&x_2^{(2)}-x_{n+1}^{(2)}&\ldots&x_2^{(n)}-x_{n+1}^{(n)}\\
\vdots&\vdots&\vdots&\vdots\\
x_n^{(1)}-x_{n+1}^{(1)}&x_n^{(2)}-x_{n+1}^{(2)}&\ldots&x_n^{(n)}-x_{n+1}^{(n)}
\end{array}\right |\ .
\end{equation}
As above the subscripts denote the points and the superscripts denote
their coordinates. 
This expression can be rewritten in a more symmetric form through the
determinant of the $(n+1)\times(n+1)$ matrix
\begin{equation}
V_n=\frac{1}{n!}
\left |\begin{array}{ccccc}
x_1^{(1)}&x_1^{(2)}&\ldots&x_1^{(n)}&1\\
x_2^{(1)}&x_2^{(2)}&\ldots&x_2^{(n)}&1\\
\vdots&\vdots&\vdots&\vdots&\vdots\\
x_n^{(1)}&x_n^{(2)}&\ldots&x_n^{(n)}&1\\
x_{n+1}^{(1)}&x_{n+1}^{(2)}&\ldots&x_{n+1}^{(n)}&1
\end{array}\right |\ .
\end{equation}
Simple manipulations show that it can be transformed in two 
different ways
\begin{equation}
V_n=\frac{(-1)^{n}}{2^nn!}\det A_n=-\frac{1}{n!}\det B_n\ ,
\end{equation}
where the $(n+2)\times (n+2)$ matrices $A_n$ and $B_n$ have the following
forms
\begin{equation}
A_n=\left (\begin{array}{ccccc}
1&0&\ldots &0&0\\
(\vec{x}_1)^2&-2x_1^{(1)}&\ldots&-2x_1^{(n)}&1\\
(\vec{x}_2)^2&-2x_2^{(1)}&\ldots&-2x_2^{(n)}&1\\
\vdots&\vdots&\vdots&\vdots&\vdots\\
(\vec{x}_n)^2&-2x_n^{(1)}&\ldots&-2x_n^{(n)}&1\\
(\vec{x}_{n+1})^2&-2x_{n+1}^{(1)}&\ldots&-2x_{n+1}^{(n)}&1
\end{array}\right )\ ,
\end{equation}
and 
\begin{equation}
B_n=\left (\begin{array}{ccccc}
0&0&\ldots &0&1\\
1&x_1^{(1)}&\ldots&x_1^{(n)}&(\vec{x}_1)^2\\
1&x_2^{(1)}&\ldots&x_2^{(n)}&(\vec{x}_2)^2\\
\vdots&\vdots&\vdots&\vdots&\vdots\\
1&x_n^{(1)}&\ldots&x_n^{(n)}&(\vec{x}_n)^2\\
1&x_{n+1}^{(1)}&\ldots&x_{n+1}^{(n)}&(\vec{x}_{n+1})^2
\end{array}\right )\ .
\end{equation}
Notice the position of the column of $1$ in $B_n$.
Therefore
\begin{equation}
V_n^2=\frac{(-1)^{n+1}}{2^n(n!)^2}\det C_n\ ,
\end{equation}
where $C_n=A_n\cdot B_n^{T}$. 

Direct calculations give the Cayler-Menger formula
\begin{equation}
V_n^2=\frac{(-1)^{n+1}}{2^n(n!)^2}
\left (\begin{array}{ccccc}
0&1&1&\ldots &1\\
1&0&D_{12}^2&\ldots&D_{1\ n+1}^2\\
1&D_{12}^2&0&\ldots&D_{2\ n+1}^2\\
\vdots&\vdots&\vdots&\vdots&\vdots\\
1&D_{1\ n}^2&\ldots&0&D_{n\ n+1}^2\\
1&D_{1\ n+1}^2&\ldots&D_{n\ n+1}^2&0
\end{array}\right )\ ,
\end{equation}
where $D_{ij}=||\vec{x}_i-\vec{x}_j||$ is the length of the edge $(i,j)$ of
the $n$-dimensional simplex.

\end{document}